\documentclass[aip, pop, twocolumn, preprintnumbers, 
amsmath, amssymb, amsfonts, floatfix, reprint]{revtex4-1}

\usepackage{adjustbox}
\usepackage{natbib}
\usepackage{graphicx}
\usepackage{dcolumn}
\usepackage{floatrow}
\usepackage{bm}
\usepackage{color}

\usepackage{multirow}
\usepackage{tabularx} 

\begin{document}

\title{Towards realistic simulations of QED cascades: non-ideal laser and electron seeding effects}

\author{Archana Sampath}
\affiliation{Max-Planck-Institut f\"ur Kernphysik, Saupfercheckweg 1, D-69117 Heidelberg, Germany}
\author{Matteo Tamburini}
\email{matteo.tamburini@mpi-hd.mpg.de}
\affiliation{Max-Planck-Institut f\"ur Kernphysik, Saupfercheckweg 1, D-69117 Heidelberg, Germany}

\date{\today}

\begin{abstract}
A number of analytical and numerical studies has been performed to investigate the onset and the development of QED cascades in the collision of two counterpropagating laser pulses as a function of the laser intensity. However, it has been recently demonstrated [M.~Tamburini \textit{et al.}, Sci. Rep. \textbf{7}, 5694 (2017)] that the onset of QED cascades is also strongly influenced by the structure of the laser pulses, such as the laser pulse waist radius. Here we investigate how QED cascades are affected by: (a)~the laser pulse duration, (b)~the presence of a relative delay for the peak of the laser pulses to reach the focus, (c)~the existence of a mismatch between the laser focal axis of the two laser pulses. 
This is especially important as, in realistic laboratory conditions, fluctuations may arise in the temporal and point stability of the lasers.
\end{abstract}


\maketitle

\section{Introduction}

High-power laser facilities such as the Extreme Light Infrastructure (ELI)~\cite{eliURL}, the Exawatt Center for Extreme Light Studies (XCELS)~\cite{xcelsURL}, the VULCAN~\cite{vulcanURL} and the Apollon-10P~\cite{apollon} projects will soon allow to attain intensities exceeding 10$^{23}$~W/cm$^2$, therefore opening up the investigation of the strong-field QED regime of laser-matter interaction in the laboratory~\cite{mourouRMP06, ehlotzkyRPP09, dipiazzaRMP12, narozhnyCP15}. 
In particular, laser-driven QED cascades initiated by seed particles will provide a prominent manifestation of the interplay between strong-field QED processes and the classical plasma dynamics. 

QED cascades are avalanche processes of hard photon emission and electron-positron pair creation driven by ultrastrong electromagnetic fields, such as two superintense colliding laser pulses. A QED cascade develops when some seed particles are violently accelerated by the electromagnetic fields therefore emitting hard photons which, in the presence of the background fields, can convert into electron-positron pairs ($e^-e^+$). 
The processes of particle acceleration with the emission of hard photons and the subsequent hard photon conversion into $e^-e^+$ pairs alternate either until all the laser pulse energy is converted into a photon-electron-positron ($\gamma e^-e^+$) plasma, or all particles are no longer inside the high field regions. Here the laser plays a dual role by transferring energy to the charged particles, which are rapidly accelerated to ultrarelativistic energies, and by providing the background field for the $\gamma$-photon emission and the non-linear Breit-Wheeler conversion of $\gamma$-photons into $e^-e^+$ pairs. The generation of QED cascades with superintense lasers allows to reproduce in the laboratory key processes occurring in astrophysical environments such as a pulsar's magnetosphere~\cite{wardleN98}, which are of critical importance for relativistic laboratory astrophysics~\cite{remingtonRMP06}. 
For this reason, QED cascades have attracted considerable attention in the last decade~\cite{bellPRL08, fedotovPRL10, elkinaPRSTAB11, nerushPRL11, bulanovjrPRL10-1, bashmakovPoP14, mironovPLA14, gelferPRA15, jirkaPRE16, grismayerPoP16, vranicPPCF17, grismayerPRE17, tamburiniSR17, artemenkoPRA17, jirkaSR17, gonoskovPRX17}. 

One of the simplest configuration for initiating a QED cascade is the collision of two counterpropagating laser pulses with seed electrons initially located at the laser focus~\cite{bellPRL08, nerushPRL11, jirkaPRE16, grismayerPoP16, grismayerPRE17, tamburiniSR17,artemenkoPRA17,jirkaSR17}. 
While it was known that the laser intensity is a crucial factor for triggering the development of a QED cascade~\cite{bellPRL08}, it has recently been shown that the laser pulse waist radius~\cite{tamburiniSR17}, the nature of the material of the target~\cite{tamburiniSR17, artemenkoPRA17}, and the target density~\cite{jirkaSR17} also play an essential role. In fact, superintense laser pulses are attained by compressing the laser pulse both temporally and spatially, which implies that strong laser field gradients are present. As a consequence, seed particles may be expelled from the focal volume due to ponderomotive effects~\cite{mulser-book} long before the laser pulse peaks reach the focus, therefore hindering or even preventing the onset of a QED cascade~\cite{tamburiniSR17}. Indeed, it has been shown that for a 20~fs laser pulse the effect of the ponderomotive expulsion of seed particles becomes increasingly important with decreasing laser pulse waist radius~\cite{tamburiniSR17}. 

The seed particle expulsion can be suppressed by employing high-$Z$ elements, whose strong nuclear electric field keeps the inner shell electrons bound to the nucleus until the laser pulse peaks are close to the focus~\cite{tamburiniSR17, artemenkoPRA17}. Another viable route to suppress seed particle expulsion is to employ solid-density targets~\cite{jirkaSR17}. In this second case, the collective fields generated by the large number of background ions results into a restoring force, which prevents some seed particles from escaping the focal volume before the laser peaks reach the focus. 
Thus, the onset of QED cascades can be either prevented with low-$Z$ gases\cite{tamburiniSR17} even at intensities around $10^{26}\text{ W/cm$^2$}$, or can be triggered at intensities approaching $10^{24}\text{ W/cm$^2$}$, which are accessible with next generation 10~PW lasers\cite{eliURL, xcelsURL, vulcanURL, apollon}, by employing high-$Z$ gases\cite{tamburiniSR17, artemenkoPRA17} or solid targets\cite{jirkaSR17}. However, we mention that there exist significant differences in the dynamics of $\gamma e^-e^+$ production in a solid target and in a tenuous gas. In fact, collective plasma effects in a solid target noticeably affect the electron dynamics and the QED cascade formation, whereas in a tenuous gas the dynamics is fully dominated by the laser fields. Indeed, plasma effects become appreciable for temporal scales of the order of the plasma period $T_{\text{pl}}=2\pi/\omega_{\text{pl}}$, where $\omega_{\text{pl}}=\sqrt{4\pi e^2 n_e /m_e}$ is the plasma frequency, $n_e$ is the electron number density, $e$ and $m_e$ are the electron charge and mass, respectively. As it will be clear in the following, since in a tenuous gas $n_{e} \lesssim 10^{-6} \, n_{\text{cr}}$ then $T_{\text{pl}} = T \sqrt{n_{\text{cr}}/n_{e}} \gg \tau$, where $\tau$ ($T$) is the laser pulse duration (period) and $n_{\text{cr}} = m_e \omega^2 / 4 \pi e^2$ is the critical plasma density. In fact, for optical lasers with frequency $\omega = 2\pi c/\lambda$ and wavelength $\lambda=0.8\text{ $\mu$m}$ we obtain $n_{\text{cr}} \approx 2\times10^{21}\text{ cm$^{-3}$}$. Thus, for a QED cascade formed in a tenuous gas where $n_e\lesssim 10^{15}\text{ cm$^{-3}$}$, throughout the laser pulse interaction the generated electrons and positrons behave like a gas under the influence of external fields, and collective effects emerge only if a dense quasi-neutral $\gamma e^-e^+$ plasma is formed.

\begin{figure}[ht]
\centering
\includegraphics[width=\linewidth]{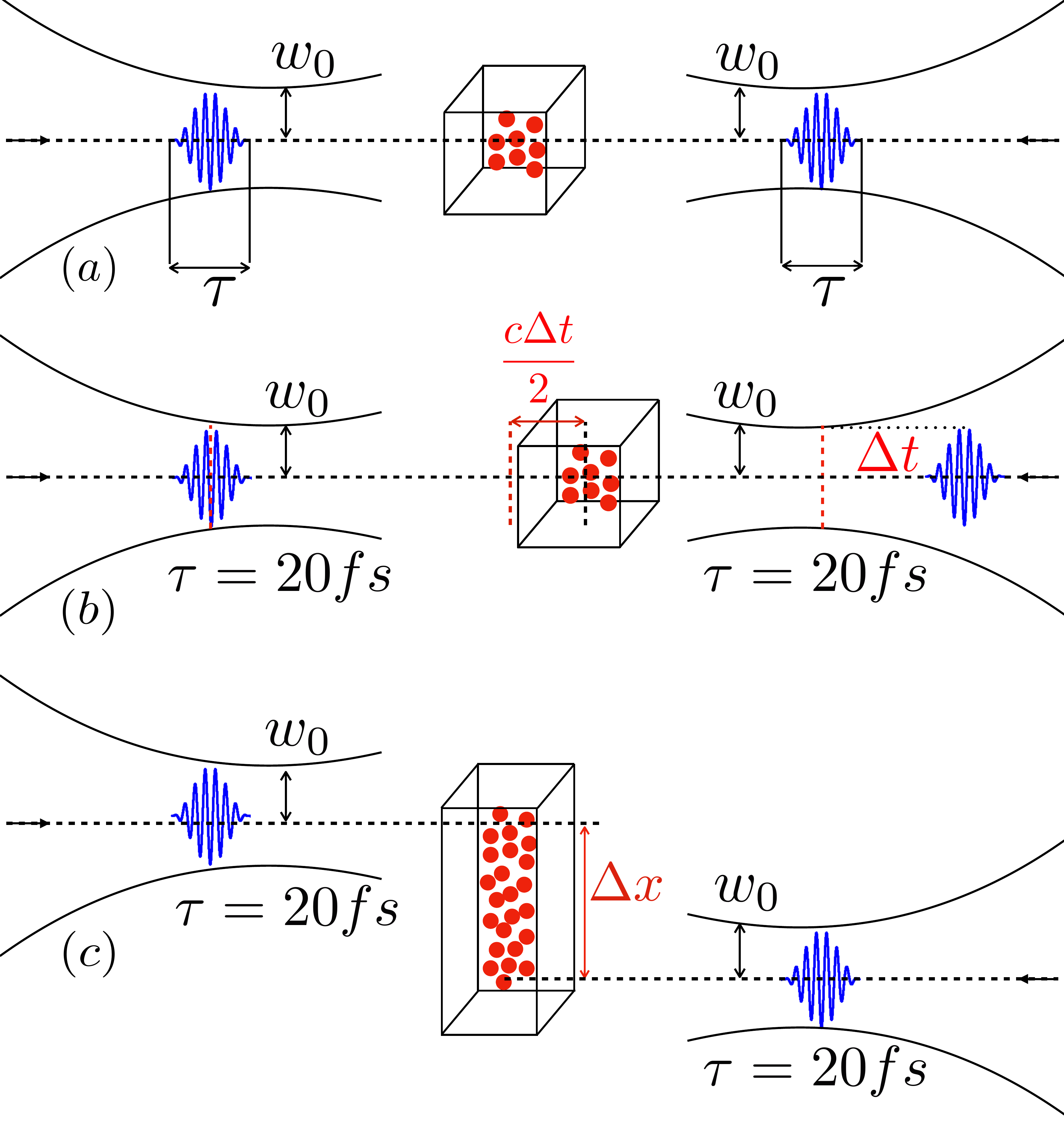}
\caption{Schematic illustration of the initial setup. (a) Effect of varying the duration $\tau$ of the two laser pulses. (b) Effect of the presence of a relative delay $\Delta t$ between the laser pulses. This implies that the two laser pulses collide at a distance $c \Delta t /2$ from the focus. (c) Effect of the presence of a misalignment $\Delta x$ between the focal axis of the laser pulses.}
\label{fig:1}
\end{figure}

Although the above-mentioned recent theoretical investigations have provided important insights that allow to better understand the factors governing the onset and the development of seeded QED cascades, there are still no studies on how the non-ideal conditions that arise during a more realistic experimental implementation may affect seeded QED cascades. 
Here we consider seeded QED cascades driven by two counterpropagating laser pulses, and explore the effects of realistic laboratory conditions such as the temporal and the pointing stability of the lasers on the cascade formation. More specifically, we investigate the effect of three parameters on the onset and the development of seeded QED cascades: (a)~the laser pulse duration $\tau$ (see Fig.~\ref{fig:1}(a)), (b)~the presence of a relative delay $\Delta t$ for the peak of the laser pulses to reach the focus (temporal stability, Fig.~\ref{fig:1}(b)), and (c)~the existence of a misalignment $\Delta x$ between the laser focal axis of the two laser pulses (pointing stability, Fig.~\ref{fig:1}(c)). 

%

\section{Setup and Method}

In our simulations, the two laser pulses are linearly polarized along the $x$-axis and propagate towards each other along the $z$-axis with their focus located at the origin. 
Both laser pulses have $\lambda=0.8\,\mu$m wavelength, $T\approx2.67$~fs period and $w_0=4\,\lambda$ waist radius. Seed electrons are randomly and uniformly distributed within a volume centered at the point of collision of the two laser pulses.
First, we have considered an initial density $n_e(0)\approx2\times10^{15}\text{ cm$^{-3}$}$ and particles distributed within a $16\lambda(x)\times16\lambda(y)\times16\lambda(z) = 4096\lambda^3$ volume. In order to check the robustness of our results, we have also performed simulations with the same parameters but either by increasing five times the initial particle density, i.e. from $n_e(0)\approx2\times10^{15}\text{ cm$^{-3}$}$ to $n_e(0)\approx10\times10^{15}\text{ cm$^{-3}$}$, or by increasing the size of the particles volume along the laser propagation direction, i.e. from $16\lambda(x)\times16\lambda(y)\times16\lambda(z) = 4096\lambda^3$ to $16\lambda(x)\times16\lambda(y)\times20\lambda(z) = 5120\lambda^3$.

A sketch of the setup for the three considered cases is displayed in Fig.~\ref{fig:1}. In the first case we investigate the effect of changing the laser pulse duration $\tau$ for two identical and perfectly synchronized laser pulses with aligned axis (see Fig.~\ref{fig:1}(a)). Three different laser pulse durations are considered: $\tau=10,\,20,\,40$~fs. In the second case the laser pulse duration is fixed to $\tau=20$~fs, but the peak of one of the two laser pulses reaches the focus with a delay $\Delta t=20,\,268$~fs after the peak of the first pulse has reached the focus (see Fig.~\ref{fig:1}(b)). Thus, the collision between the peak of the two laser pulses occurs at $z^*=c\Delta t /2$ and the center of the considered volume of particles is consequently shifted to $z^*$. Finally, in the third case the laser pulse duration is fixed to $\tau=20$~fs, but the focal spot of one of the lasers is shifted from the origin to $\Delta x$ along the $x$-axis (see Fig.~\ref{fig:1}(c)). Note that in the presence of a transverse misalignment $\Delta x$ the considered volume of particles is correspondingly increased either to $(16\lambda + \Delta x) (x) \times 16\lambda(y) \times 16\lambda (z)$ or to $(16\lambda + \Delta x) (x) \times 16\lambda(y) \times 20\lambda (z)$. In all the setups, calculations are performed either for a power $P=200$~PW or for a power $P=300$~PW per laser pulse. Seed electrons are assumed to originate from low density hydrogen gas, which is the main contaminant in current ultrahigh vacuum technology~\cite{fremereyV99}. 

In order to investigate the full scale evolution of seeded QED cascades, i.e. from the ionization of the hydrogen gas in the wings of the laser pulses to the emission of high-energy photons and the formation of $e^-e^+$ pairs, we performed three-dimensional Monte~Carlo simulations with the initial position $z_0$ of the laser peaks chosen such that they collide at the center of the particle volume after 70~$T$ and 150~$T$ for $\tau\leq20$~fs and $\tau>20$~fs, respectively. This ensures that, initially, the fields at the focus are much smaller than the field required to ionize hydrogen atoms. In our simulations the ionization of hydrogen gas is accounted for by a simple model, where electrons go into the continuum, i.e. are free to move, when the electric field at the electron positions exceeds the hydrogen over-the-barrier ionization field $E_{\text{H}}$. For high-$Z$ atoms more accurate and complete models are needed to account for the ionization of electrons belonging to different atomic shells~\cite{artemenkoPRA17}. For hydrogen in the ground state a quantum mechanical calculation gives~\cite{mulser-book}:
\begin{equation}
E_{\text{H}} = \frac{(\sqrt{2}-1)}{2\sqrt{2}}\frac{m_e^{2}|e|^{5}}{\hbar^{4}},
\end{equation}
which corresponds to an intensity of the order of $10^{14}$~W/cm$^2$, whereas the peak laser intensities considered here are greater than $10^{24}$~W/cm$^2$. Thus, electrons go into the continuum long before the peak of the laser pulses reach the focal spot. This implies that an accurate temporal description of the laser fields even in the wings of the laser pulses is required. For the above reasons, in our simulations the laser pulse has a sech$^2$ temporal intensity profile, which was shown to bear a close similarity to shape of short laser pulses currently produced in the laboratory~\cite{asakiOL93, lazaridisOL95}. In addition, from the theoretical point of view, a sech$^2$ temporal intensity profile provides an accurate description of short and tightly focused laser pulses in their entire temporal domain~\cite{mcdonaldURL}. 

In the code, the dynamics of electrons and positrons between subsequent photon emissions is calculated with the Lorentz force, while the stochastic processes of photon emission and pair creation is modeled using a Monte Carlo algorithm described in Ref.~\cite{tamburiniSR17}. The code employs an adaptive fourth-order Runge-Kutta integrator with time step $\delta t$ chosen such that two conditions must be simultaneously fulfilled, namely $\delta t < T / (200 \pi)$ and also $\delta t < T / (10 a_{0})$, where $a_{0} = e E_{0} / m_e \omega c$ is the normalized field amplitude, $T$ is the laser period, and $E_{0}$ is the maximum of the local value of the electric and magnetic field at the particle's position. This is important because for larger electromagnetic fields a smaller time step is needed in order to accurately model both the complex particle dynamics and stochastic processes such as photon emission and pair creation. In fact, for example, the photon emission probability per unit time is proportional\cite{Baier-book} to $a_{0}$. Note that for the intensities considered in our manuscript $a_{0}$ can exceed $10^3$ such that a very small time step is needed. In all our simulations we have employed a fully three-dimensional description of the laser pulse fields with terms up to the fifth order in the diffraction angle~\cite{salaminAPB07} $\epsilon = \lambda / \pi w_0$. Thus, unlike particle-in-cell (PIC) codes, no minimal spatial step is present because no spatial grid is used. In addition, in our simulations there is a one-to-one correspondence between physical particles and numerical particles, this means that within the considered volume each computational particle represents a real particle. No particle merging scheme was used to avoid to introduce possible numerical effects. Hence, the only source of fluctuations or noise arises from the fact that photon emission and pair creation are intrinsically stochastic processes. This kind of statistical fluctuations is of physical origin and not related to numerical effects. We stress that collective effects remain negligible for the spatio-temporal scales relevant in our scenario due to the low initial density. Indeed, $T_{\text{pl}} = T \sqrt{n_{\text{cr}}/n_{e^-}(0)}\approx10^3\,T$ since with our parameters $n_{e^-}(0) \approx 10^{-6} \, n_{\text{cr}}$, whereas $T_{\text{pl}}\gg\tau\approx 4\,T$-$15\,T$. Self-generated fields may become important if the electron/positron density is comparable to the critical plasma density $n_{\text{cr}} \approx 2\times10^{21} \text{ cm$^{-3}$}$ for example because there are more initial particles to seed a cascade, such as in a solid target, or because the number of created electron-positron pairs increases the electron-positron density up to $n_{\text{cr}}$. However, none of the above-mentioned conditions occur in our simulations. Note that when the normalized laser field amplitude $a_0$ becomes much larger than unity, then the critical plasma density is increased from $n_{\text{cr}}$ to the relativistic critical plasma density $a_0 \, n_{\text{cr}}$. This implies that collective plasma effects are more suppressed when $a_0\gg1$ and become important for electron/positron densities $a_0$ times larger than $n_{\text{cr}}$, or for temporal scales comparable to $T_{\text{pl}} \approx T \sqrt{a_0\,n_{\text{cr}}/n_{e^-}(0)}$. Regarding the possible effects of ions, since ionization occurs at much lower intensities than the considered laser peak intensities, and since the ion charge to mass ratio is at least $2 \times 10^{3}$ times smaller than the electron charge to mass ration, ions do not move significantly during the ionization phase. After ionization, their comparatively low number and the short duration of the considered laser pulses strongly suppresses their possible influence on the onset or development of a QED.

\section{Results}

Strong-field QED processes are controlled by the electron/photon quantum parameter~\cite{Baier-book, ritusJSLR85} $\chi_{e/\gamma}=\sqrt{|(F_{\mu\nu}p^{\nu}_{e/\gamma})^2|}/F_{\text{cr}}m_e c$, where $F_{\mu\nu}$ is the field tensor, $p^{\nu}_{e/\gamma}$ the electron/photon four-momentum, and $F_{\text{cr}} = m_e^2 c^3 / |e| \hbar \approx 1.3\times10^{16}\text{ V/cm} \approx 4.4 \times 10^{13}\text{ G}$ is the QED critical field. The quantum parameter $\chi_{e/\gamma}$ determines the probability of the two basic strong-field QED processes: photon emission by electrons and positrons in a strong electromagnetic field, and the conversion of a high-energy photon into an $e^-e^+$ pair via the non-linear Breit-Wheeler process. In fact,
for $\chi_{e}\ll 1$ the typical energy of the emitted photons is $\varepsilon_\gamma\approx\chi_e \varepsilon_e$, where $\varepsilon_e$ is the electron energy, while for $\chi_{\gamma}\ll 1$ the probability of photon conversion into an $e^{-}e^{+}$ pair is exponentially suppressed~\cite{Baier-book, ritusJSLR85} as $e^{-8/3\chi_\gamma}$. Hence, single photon emission recoil is important as $\chi_e\gtrsim 1$, and pair creation is significant only when $\chi_\gamma\gtrsim 1$. Thus, in order to create an avalanche of $e^-e^+$ pairs, i.e. the onset and the development of a QED cascade, particles must attain  $\chi_{e/\gamma}\gtrsim 1$ for sufficiently long time, such that there is an exponential growth in the number of $e^{-}e^{+}$ pairs with the creation on average of at least one pair per initial electron~\cite{fedotovPRL10, gelferPRA15}. In the following, other scenarios where $e^{-}e^{+}$ pairs are produced but less than one pair per initial electron is produced are termed ``$\gamma e^{-}e^{+}$ gases''~\cite{tamburiniSR17}. 

\begin{table*}[!th]
\centering
\begin{tabular}{|c|c|c|c|c|c|c|c|c|c|c|}
\hline
\multirow{2}{*}{$P$ (PW)} & \multirow{2}{*}{$\tau$ (fs)} & \multicolumn{3}{c|}{\begin{tabular}[c]{@{}c@{}}$n_{e^-}(0)\approx 2 \times 10^{15}$cm$^{-3}$\\ $16\lambda(x) \times 16\lambda(y) \times 16\lambda (z)$\end{tabular}} & \multicolumn{3}{c|}{\begin{tabular}[c]{@{}c@{}}$n_{e^-}(0)\approx 10 \times 10^{15}$cm$^{-3}$\\ $16\lambda(x) \times 16\lambda(y) \times 16\lambda (z)$\end{tabular}} & \multicolumn{3}{c|}{\begin{tabular}[c]{@{}c@{}}$n_{e^-}(0)\approx 2 \times 10^{15}$cm$^{-3}$\\ $16\lambda (x) \times 16\lambda(y) \times 20\lambda (z)$ \end{tabular}} \\ \cline{3-11}  
                          &                              & Status & $N_{e^-e^+}$         & $N_{\gamma}$         & Status & $N_{e^-e^+}$ & $N_{\gamma}$ & Status & $N_{e^-e^+}$ & $N_{\gamma}$ \\ \hline
\multirow{3}{*}{200}      & 10                           & G      & 4.70 $\times$ 10$^4$ & 4.51 $\times$ 10$^6$ & G      &   2.43 $\times$ 10$^5$     & 2.32 $\times$ 10$^7$ & G   &  5.10 $\times$ 10$^4$     & 4.84 $\times$ 10$^6$  \\  
                          & 20                           & G      & 1.97 $\times$ 10$^4$ & 2.12 $\times$ 10$^6$ & G      & 1.69 $\times$ 10$^5$     & 1.83 $\times$ 10$^7$ & G      & 2.69 $\times$ 10$^4$     & 2.90 $\times$ 10$^6$  \\  
                          & 40                           & N      & 0                    & 0                    & N      & 0                 & 0 & N      &      0        &           0   \\ \hline
\multirow{3}{*}{300}      & 10                           & G      & 1.03 $\times$ 10$^6$ & 6.98 $\times$ 10$^7$ & G      & 4.95 $\times$ 10$^6$     & 3.36 $\times$ 10$^8$ & G      &  1.05 $\times$ 10$^6$     & 7.15 $\times$ 10$^7$              \\ 
                          & 20                           & C      & 1.75 $\times$ 10$^7$ & 1.76 $\times$ 10$^9$   & C     & 7.13 $\times$ 10$^{7}$			       &  6.79  $\times$ 10$^{9}$ & C      &       1.28 $\times$ 10$^{7}$ 			       & 1.27  $\times$ 10$^{9}$              \\
                          & 40                           & N      & 0                    & 1                    & N      &      0        &     0         & N      &        0      &           0   \\ \hline
\end{tabular}
\caption{The number of $e^-e^+$ pairs $N_{e^-e^+}$ and photons $N_{\gamma}$ with energy larger than 2.5~MeV produced as a function of the laser pulse duration $\tau$ in the collision of two laser pulses each with waist radius $w_0=4\lambda$ and power $P$ of either 200 or 300~PW. Results are reported for three different sets of simulation parameters, where either the particle density or the particle volume was changed.
Symbols are as follows: N = No $e^{-}e^{+}$ pairs, G = $e^{-}e^{+}$ gas, C = $\gamma e^{-}e^{+}$ cascade.}
\label{tab:1}
\end{table*}
\begin{figure*}[t]
\centering
\includegraphics[width=\linewidth]{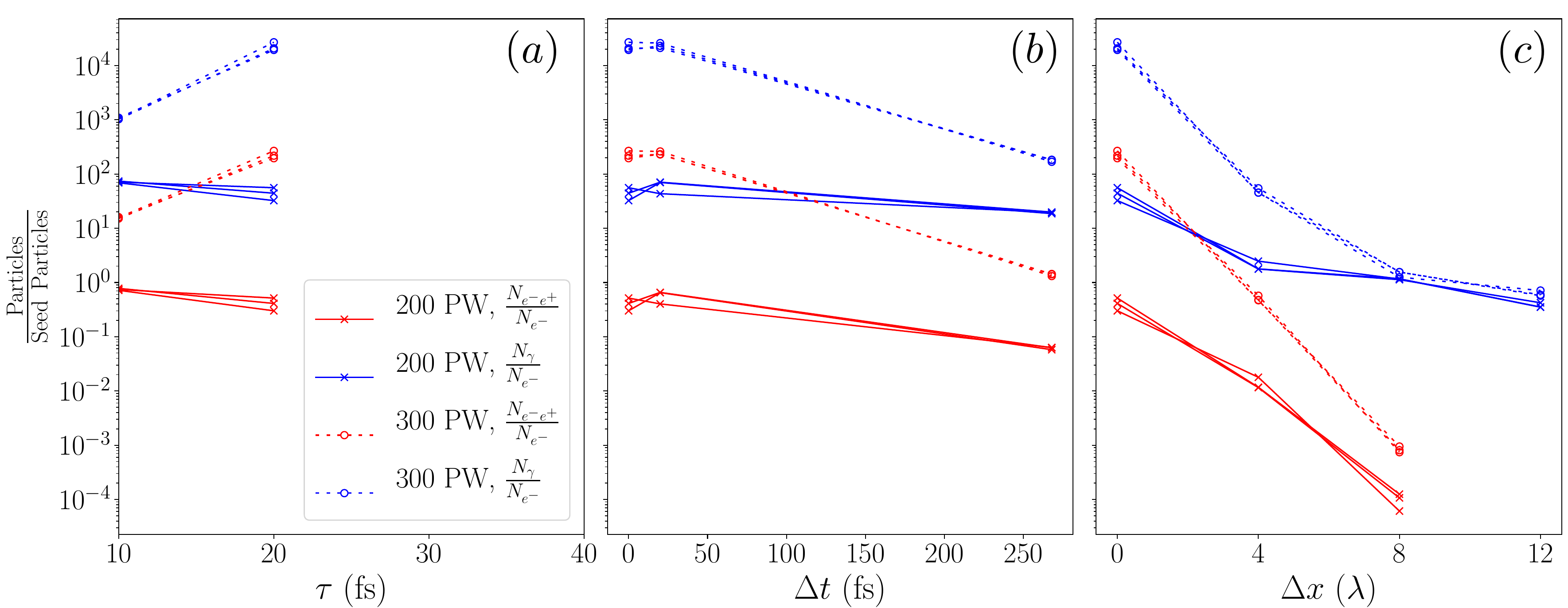}
\caption{The number of created $e^-e^+$ pairs $N_{e^-e^+}$ (photons $N_\gamma$ with energy greater than 2.5 MeV) divided by the initial number of seed electrons $N_{e^-}$ in a $w_0^3$ volume in the collision of two laser pulses each with waist radius $w_0$ = 4$\lambda$ and power either of 200 or of 300~PW. Note that for the cases where no particles are created no data is displayed. (a)  The number of created particles as a function of the laser pulse duration $\tau$. (b) and (c) display the number of created particles for two 20~fs laser pulses in the presence of a relative delay $\Delta t$ and a transverse misalignment $\Delta x$, respectively.}
\label{fig:tab}
\end{figure*}
\begin{figure*}[t]
\centering
\includegraphics[width=\linewidth]{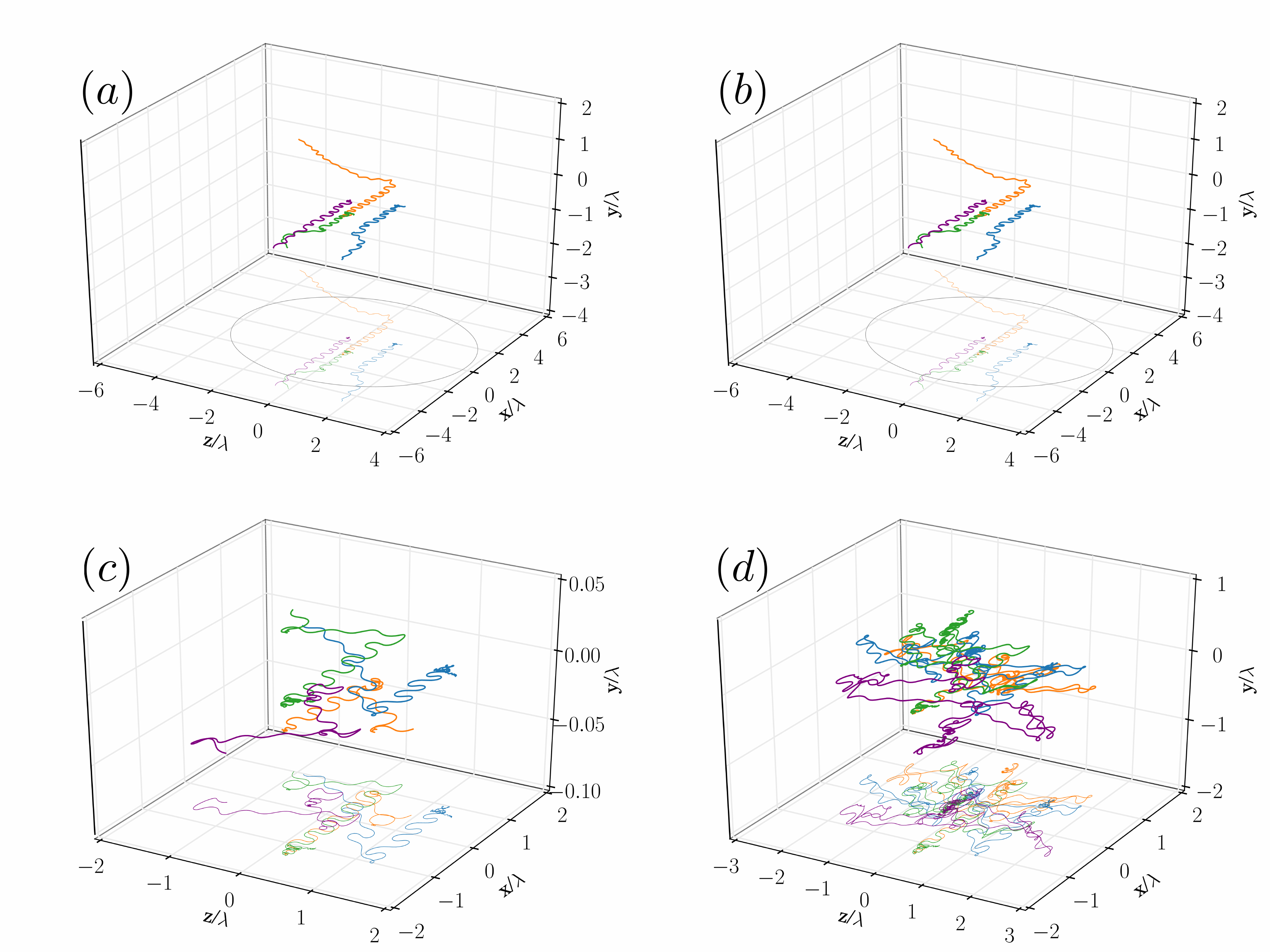}
\caption{The trajectories of four seed electrons driven by two 300~PW tightly-focused ($w_0 = 4\lambda$) counter-propagating laser pulses. The projection of the trajectories on the $xz$ plane is also reported, $z$ ($x$) being the laser propagation (polarization) direction. (a) The two laser pulses have 40~fs duration and the displayed trajectories correspond to the temporal interval $70 < t < 92\,T$. At $t = 92\,T$ the intensity at the focus is $I \approx 9.3 \times 10^{19}~\text{ W/cm$^2$}$. (b) Same as in (a) but for the larger temporal interval $70 < t < 140\,T$. Here only the part of the trajectory within the radius $r = \sqrt{x^2+y^2+z^2} \leq 5 \lambda$ is displayed, showing that no electron returns inside the focal region. (c) The two laser pulses have 20~fs duration and the displayed trajectories correspond to the temporal interval $0 < t < 46\,T$. At $t = 46\,T$ the intensity at the focus is $I \approx 9.3 \times 10^{19}~\text{ W/cm$^2$}$. (d) Same as in (c) but for the whole temporal interval until the intensity at the focus reaches its maximal value $I \approx 1.9 \times 10^{24}~\text{ W/cm$^2$}$.}
\label{fig:2}
\end{figure*}
\begin{figure}[t]
\centering
\includegraphics[width=\linewidth]{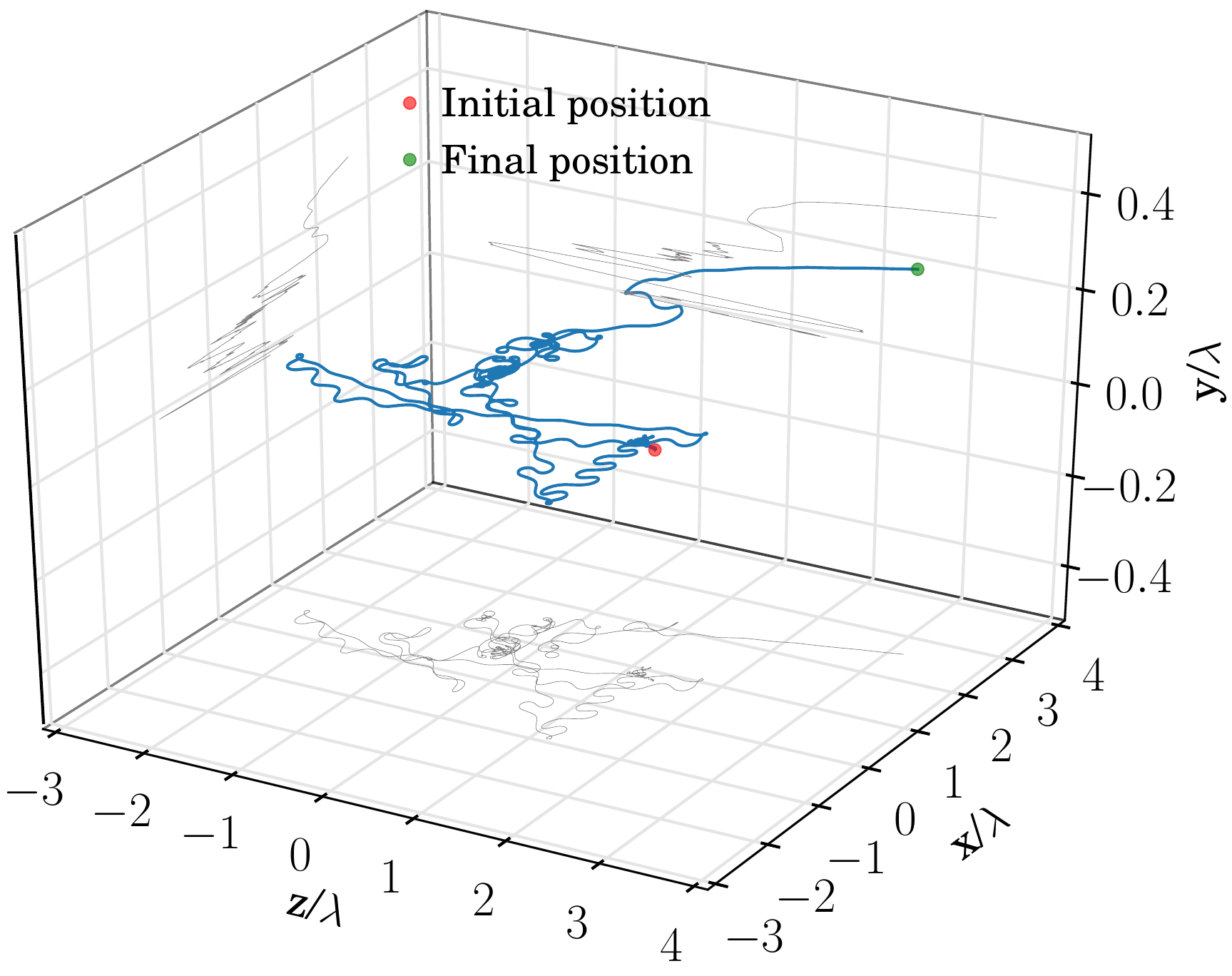}
\caption{The trajectory of a single seed electron driven by two 300~PW and 20~fs tightly-focused ($w_0 = 4\lambda$) counter-propagating laser pulses. The maximal laser intensity is $I \approx 1.9 \times 10^{24}~\text{ W/cm$^2$}$ and is reached at $t=70\,T$ while the displayed electron trajectories correspond to the temporal interval $0 < t < 85.1\,T$. The projection of the trajectory on the $xz$, $xy$ and $yz$ is also reported.}
\label{fig:3}
\end{figure}
\begin{figure*}[!h]
\centering
\includegraphics[width=\linewidth]{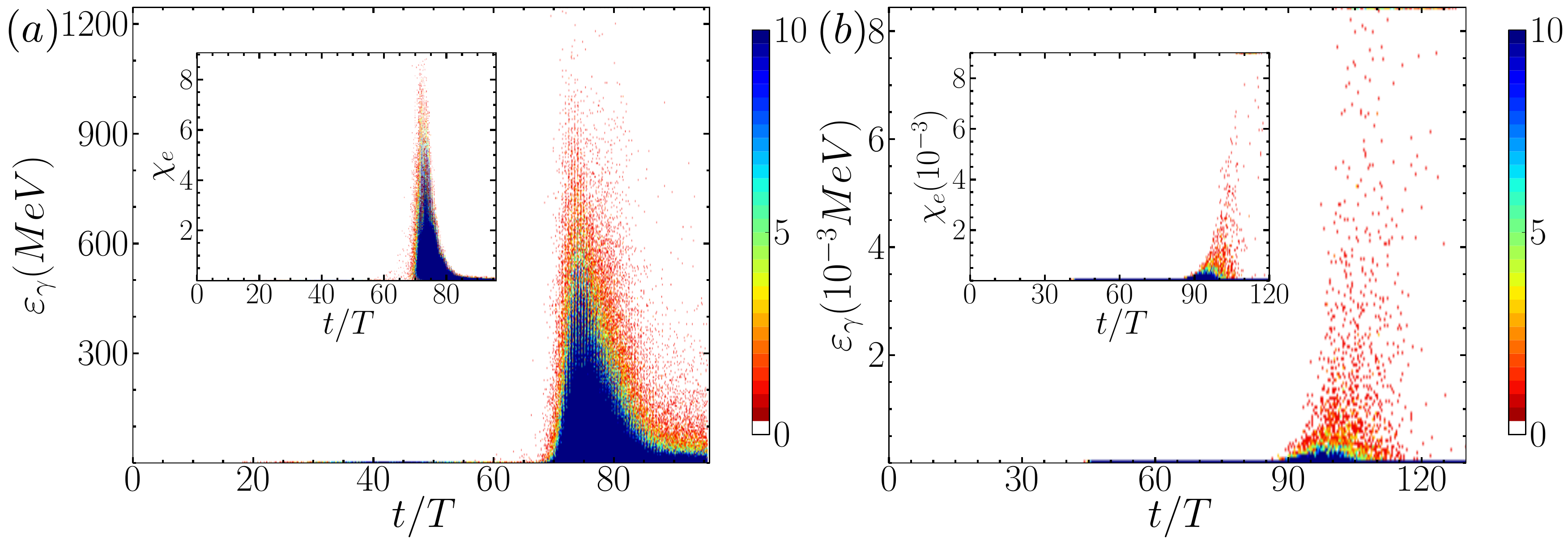}
\caption{The emitted photon energy $\varepsilon_\gamma$ and the electron quantum parameter $\chi_e$ (inset) at each photon emission event as function of time $t$. Each laser pulse has $P=300$~PW and $w_0=4\,\lambda$, i.e. intensity $I \approx 1.9 \times 10^{24}\text{ W/cm$^2$}$. (a) Laser pulse duration $\tau=20$~fs and the two laser pulse peak are located at $z_0=\pm70\,\lambda$, initially. (b) Laser pulse duration $\tau=40$~fs and the two laser pulse peak are located at $z_0=\pm150\,\lambda$, initially. The color bar levels correspond to the number of events (black indicates at least 10 events).}
\label{fig:4}
\end{figure*}
\begin{table*}[!h]
\centering
\begin{tabular}{|c|c|c|c|c|c|c|c|c|c|c|}
\hline
\multirow{2}{*}{$P$ (PW)} & \multirow{2}{*}{$\Delta t$ (fs)} & \multicolumn{3}{c|}{\begin{tabular}[c]{@{}c@{}}$n_{e^-}(0)\approx 2 \times 10^{15}$cm$^{-3}$\\ $16\lambda(x) \times 16\lambda(y) \times 16\lambda (z)$\end{tabular}} & \multicolumn{3}{c|}{\begin{tabular}[c]{@{}c@{}}$n_{e^-}(0)\approx 10 \times 10^{15}$cm$^{-3}$\\ $16\lambda(x) \times 16\lambda(y) \times 16\lambda (z)$\end{tabular}} & \multicolumn{3}{c|}{\begin{tabular}[c]{@{}c@{}}$n_{e^-}(0)\approx 2 \times 10^{15}$cm$^{-3}$\\ $16\lambda (x) \times 16\lambda(y) \times 20\lambda (z)$ \end{tabular}} \\ \cline{3-11}  
                          &                              & Status                                            & $N_{e^-e^+}$                                            & $N_{\gamma}$                                           & Status                                            & $N_{e^-e^+}$                                            & $N_{\gamma}$                                           & Status                                            & $N_{e^-e^+}$                                            & $N_{\gamma}$                                           \\ \hline
\multirow{3}{*}{200}      &   0                      & G      & 1.97 $\times$ 10$^4$   & 2.12 $\times$ 10$^6$                                                                                                              & G      & 1.69 $\times$ 10$^5$   & 1.83 $\times$ 10$^7$                                                         & G      & 2.69 $\times$ 10$^4$   & 2.90 $\times$ 10$^6$                                                       \\  
                          & 20                     & G      & 4.26 $\times$ 10$^4$   & 4.57 $\times$ 10$^6$                                                       & G      & 1.32 $\times$ 10$^5$   & 1.42 $\times$ 10$^7$                                                   & G      & 4.30 $\times$ 10$^4$   & 4.63 $\times$ 10$^6$                                                        \\ 
                          & 268                           &          G                                         &                          3.78 $\times$ 10$^{3}$         &         1.22 $\times$ 10$^{6}$                                                  &          G                         &                  2.07 $\times$ 10$^{4}$                 &       6.42 $\times$ 10$^{6}$                                      &                 G                                       &                 4.10 $\times$ 10$^{3}$                                         &           1.30 $\times$ 10$^6$                                                                                                    \\ \hline
\multirow{3}{*}{300}       & 0                      & C      & 1.75 $\times$ 10$^7$   & 1.76 $\times$ 10$^9$                                                        & C     & 7.13 $\times$ 10$^{7}$            &  6.79  $\times$ 10$^{9}$                                                          & C     &  1.28 $\times$ 10$^{7}$ 			       & 1.27  $\times$ 10$^{9}$                                                        \\ 
                          & 20                     & C      & 1.71 $\times$ 10$^7$   & 1.70 $\times$ 10$^9$                                                        & C     &  7.52 $\times$ 10$^{7}$            & 6.91  $\times$ 10$^{9}$                                                             & C     &  1.52 $\times$ 10$^7$    & 1.51 $\times$ 10$^9$                                                       \\ 
                          & 268                        &           G                                        &                   8.60 $\times$ 10$^4$               &       1.10 $\times$ 10$^7$                                                 &                       G                              &       4.70 $\times$ 10$^5$                                                  &                     6.00 $\times$ 10$^7$             &           G                                        &                     9.29 $\times$ 10$^4$             &        1.19 $\times$ 10$^7$                                                \\ \hline \hline
\multirow{2}{*}{$P$ (PW)} & \multirow{2}{*}{$\Delta x$ ($\lambda$)} & \multicolumn{3}{c|}{\begin{tabular}[c]{@{}c@{}}$n_{e^-}(0)\approx 2 \times 10^{15}$cm$^{-3}$\\ $(16\lambda + \Delta x) (x)  \times 16\lambda(y) \times 16\lambda (z)$\end{tabular}} & \multicolumn{3}{c|}{\begin{tabular}[c]{@{}c@{}}$n_{e^-}(0)\approx 10 \times 10^{15}$cm$^{-3}$\\ $(16\lambda + \Delta x) (x)  \times 16\lambda(y) \times 16\lambda (z)$\end{tabular}} & \multicolumn{3}{c|}{\begin{tabular}[c]{@{}c@{}}$n_{e^-}(0)\approx 2 \times 10^{15}$cm$^{-3}$\\$(16\lambda + \Delta x) (x) \times 16\lambda(y) \times 20\lambda (z)$ \end{tabular}} \\ \cline{3-11}   
                          &      & Status                                            & $N_{e^-e^+}$                                            & $N_{\gamma}$                                           & Status                                            & $N_{e^-e^+}$                                            & $N_{\gamma}$                                           & Status                                            & $N_{e^-e^+}$                                            & $N_{\gamma}$                                           \\ \hline
\multirow{4}{*}{200}      & 0                      & G      & 1.97 $\times$ 10$^4$   &          2.12 $\times$ 10$^6$                                                        & G      & 1.69 $\times$ 10$^5$   & 1.83 $\times$ 10$^7$                                                            & G      & 2.69 $\times$ 10$^4$   & 2.90 $\times$ 10$^6$                                                        \\ 
                           & 4                      & G      & 1.17 $\times$ 10$^3$    &         1.62 $\times$ 10$^5$                                                          & G      & 3.83 $\times$ 10$^3$   & 5.83 $\times$ 10$^5$                                                     & G      & 7.56 $\times$ 10$^2$   & 1.16  $\times$ 10$^5$                                                         \\ 
                         & 8                      & G      & 4               &          7.60 $\times$ 10$^4$                                                         & G      & 4.10 $\times$ 10$^1$   & 3.83 $\times$ 10$^5$                                                       & G      & 7   & 7.38 $\times$ 10$^4$                                                        \\ 
                          & 12                     & N      & 0               &          2.31 $\times$ 10$^4$                                                         & N      & 0               		& 1.15 $\times$ 10$^5$                                                          & N      & 0               		& 2.77 $\times$ 10$^4$                                                        \\ \hline
\multirow{4}{*}{300}      & 0                      & C      & 1.75 $\times$ 10$^7$   &          1.76 $\times$ 10$^9$                                                          & C     & 7.13 $\times$ 10$^{7}$			       &  6.79  $\times$ 10$^{9}$                                      & C      &  1.28 $\times$ 10$^{7}$ 			       & 1.27  $\times$ 10$^{9}$                                                        \\  
                        & 4                      & G      & 3.12 $\times$ 10$^4$    &         2.98 $\times$ 10$^6$                                                         & G      & 1.87 $\times$ 10$^5$   & 1.79 $\times$ 10$^7$                                                          & G      & 3.08 $\times$ 10$^4$   & 2.97 $\times$ 10$^6$                                                        \\ 
                           & 8                      & G      & 6.20 $\times$ 10$^1$   &          1.02 $\times$ 10$^5$                                                          & G      & 2.43 $\times$ 10$^2$   & 5.10 $\times$ 10$^5$                                                      & G      & 5.3 $\times$ 10$^1$   & 8.21 $\times$ 10$^4$                                                      \\ 
                           & 12                     & N      & 0               &          3.74 $\times$ 10$^4$                                                          & N      & 0               		& 1.94 $\times$ 10$^5$                                                       & N      & 0              		& 4.66 $\times$ 10$^4$                                                         \\ \hline
\end{tabular}
\caption{The number of $e^-e^+$ pairs $N_{e^-e^+}$ and photons $N_{\gamma}$ with energy larger than 2.5~MeV produced as a function either of the relative delay $\Delta t$ or of the transverse misalignment $\Delta x$ between two 20~fs laser pulses each with waist radius $w_0=4\lambda$ and power $P$ of either 200 or 300~PW. Results are reported for three different sets of simulation parameters, where either the particle density or the particle volume was changed. Symbols are as follows: N = No $e^{-}e^{+}$ pairs, G = $e^{-}e^{+}$ gas, C = $\gamma e^{-}e^{+}$ cascade.}
\label{tab:2}
\end{table*}

We first discuss the effect of the laser pulse duration on the onset of QED cascades. It has been already pointed out that the spatio-temporal structure of the laser pulses is of critical importance for the onset of QED cascades~\cite{tamburiniSR17}. In fact, the laser pulse structure determines both when electrons go into the continuum due to atom ionization, and it also determines whether particles are eventually expelled out of the focal volume before the fields become sufficiently intense to trigger a QED cascade. In fact, on the one hand the spatial gradients unavoidably associated to the tight focusing of intense laser pulses control the strength of ponderomotive effects, which drive particles out of the focal region~\cite{tamburiniSR17}. 
For instance, for a single laser pulse described by the vector potential $\mathbf{A}(\mathbf{r},t)$, the secular $e^-e^+$ dynamics is known to be determined by the ponderomotive force~\cite{mulser-book}:
\begin{equation}
\mathbf{f}_p = -m_e c^2\boldsymbol{\nabla}(1+\langle \mathbf{a}^2 \rangle)^{1/2},
\end{equation}
where $\mathbf{a}=e \mathbf{A} / m_e c^2$ is the normalized vector potential. 
On the other hand, the temporal gradients control the time particles can spend inside the focal volume before being either pondermotively expelled, which prevents the formation of a seeded QED cascade, or attaining $\chi_{e/\gamma}\gtrsim 1$, which may trigger a QED cascade. 
Thus, the expulsion of seed electrons before the laser field amplitude reaches values sufficiently large to trigger the creation of electron-positron pairs is determined both by the strength of ponderomotive effects, which is controlled by the spatial gradients, and by the time available to ponderomotive effect to operate, which is controlled by the temporal gradients. For a Gaussian laser beam with sech$^2$ intensity envelope the spatial and the temporal field gradients are determined by $w_0$ and by $\tau$, respectively.
Note that the ponderomotive effect leading to seed particle expulsion is especially important in the temporal window ranging from ionization ($a_0\sim 0.01$ for hydrogen) to efficient gamma photon emission ($a_0\sim 100$). The temporal gradients depend on $\tau$, for larger values of $\tau$ the ponderomotive force has more time to expel seed electrons before $a_0\gtrsim 300$ and pair creation probability becomes appreciable. If some seed electrons are still present in the focal region when $a_0\gtrsim 300$, then the QED cascade develops predominantly in a temporal region of the order of $\tau$ around the peak of the laser pulse. Thus, on the one hand a larger $\tau$ would potentially imply a longer duration of the QED cascade but, on the other hand, a larger $\tau$ reduces the number of seed particles in the focal volume. This is similar to Ref.~\cite{tamburiniSR17} for fixed laser pulse duration but varying the laser waist radius $w_0$. For fixed laser power, on the one hand a smaller $w_0$ the laser intensity is higher and the cascade growth rate $\Gamma$ is potentially larger. On the other hand, a smaller $w_0$ increases the field gradients and consequently the strength of ponderomotive effects therefore reducing the number of seed particles in the focal volume. This implies that for fixed laser pulse power and duration there exists an optimal waist radius or, vice versa, for fixed laser power and waist radius there exists an optimal laser pulse duration.

Table~\ref{tab:1} and Fig.~\ref{fig:tab}(a) summarize the results obtained with fixed waist radius $w_0=4\,\lambda$ but different laser pulse duration $\tau=10,\,20,\,40$~fs. Table~\ref{tab:1} shows that a relatively long 40~fs laser pulse provides enough time for the electrons to gain momentum in the wings of the pulse and get expelled out of the focal volume before the laser peaks reach the focus. 
Indeed, the increase of the laser pulses duration from 20 to 40~fs results into the expulsion of all seed electrons from the focal volume, therefore drastically suppressing the number of generated $e^-e^+$ pairs (see Tab.~\ref{tab:1}). 
The different maxima in the electron-positron pair yield reported in Tab.~\ref{tab:1} can be explained by recalling that the final number of electron-positron pairs depends on the initial number of particles actually seeding the cascade $N_{0,e^{-}}$, on the average growth rate $\langle\Gamma\rangle$, and on the duration $\tau$ of the cascade itself. In fact, the final positron yield $N_{e^{+}}(\infty)$ can be parametrized\cite{bashmakovPoP14, tamburiniSR17} as $N_{e^{+}}(\infty) \approx N_{0,e^{-}} [e^{\langle\Gamma\rangle \tau} - 1] / 2$. Thus, there are three quantities that determine the final number of pairs, namely $N_{0,e^{-}}$, $\langle\Gamma\rangle$, and $ \tau$. 
The number of seed electrons actually seeding the cascade can be parametrized as: $N_{0,e^{-}} = N_{e^-}(t_i)  e^{-(t_c-t_i) / \tau_0}$, where $t_i$ is the time of ionization, i.e. when $a_0\sim0.01$ for hydrogen, $t_c$ is the time when the probability of $e^-e^+$ production becomes appreciable, i.e. approximately when $a_0\gtrsim 300$, and $\tau_0$ is the effective average escape rate due to ponderomotive effects, which mainly depends on $w_0$.
Note that $(t_c-t_i)$ increases with increasing $\tau$, but it is larger than $\tau$. In fact, by using the expression for the temporal field envelope of the laser pulse we find:
\begin{equation} \label{time}
(t_c - t_i) = \frac{\tau}{2 \cosh^{-1} (\sqrt{2})} \log \left(\frac{a_{0_c} \left(a_0 + \sqrt{a_0^2 - a_{0_i}^2}\right)} {a_{0_i} \left(a_0 + \sqrt{a_0^2 - a_{0_c}^2}\right)} \right).
\end{equation}
Assuming the normalized field amplitude for the onset of a QED cascade $a_{0_c} \approx 300$, for hydrogen ionization $a_{0_i} \approx 0.01$ and peak value $a_0 \approx 930$, which corresponds to 300~PW power and $w_0=4\lambda$, Eq.~(\ref{time}) gives $(t_c - t_i) \approx 5.9 \, \tau$, i.e. $(t_c - t_i)=118\text{ (236) fs}$ for a 20~(40)~fs laser pulse. Regarding $\tau_0$, a simple estimate as $\tau_0 \sim w_0/c$ gives $\tau_0 \approx 11\text{ fs}$ for $w_0=4\lambda$, but simulations indicate a larger value $\tau_0\approx17\text{ fs}$. Thus, 
the final positron yield $N_{e^{+}}(\infty)$ can exceed the initial number of seed particles if: (i) $N_{0,e^{-}} \geq 1$ such that seed particles are present in the focal volume, and (ii) the average growth rate $\langle\Gamma\rangle$ and the time available for a QED cascade to develop $\tau$ are sufficiently large to compensate for the decrease in the number of seed particles, i.e. $\langle\Gamma\rangle \tau > (t_c-t_i) / \tau_0$.

For fixed laser pulse power and waist radius $w_0$, while $N_{0,e^{-}}$ decreases with increasing laser pulse duration $\tau$ and reaches zero for $\tau=40$~fs, the product $\langle\Gamma\rangle \tau$ increases with increasing laser pulse duration. 
%
As a result, for 200 PW laser pulses a higher $e^+e^-$ yield is obtained for a 10~fs rather than for a 20~fs laser pulse because the decrease in  $N_{0,e^{-}}$ for larger $\tau$ prevails over the increase of $\langle\Gamma\rangle \tau$. By contrast, the increase in the laser field amplitude in the 300~PW case is such that $\langle\Gamma\rangle$ increases and the larger value of $\langle\Gamma\rangle \tau$ prevails over the decrease of  $N_{0,e^{-}}$ for a 20~fs pulse. 

In order to determine $N_{0,e^{-}}$, in our simulations each particle initially present is labeled with a unique identifier, such that this identifier is inherited by all the new generations of particles generated during the simulation. The identifier allowed us to single out the initial particles that originated a large number of electron positron-pairs. Interestingly, even for the case of shorter pulses that do trigger a QED cascade, most of the initial particles are prematurely expelled out of the focal volume, and the final yield of electron-positron pairs originates from a relatively small fraction of the particles initially located in the focal volume. We have therefore tracked the dynamics at each time step of a sample of ten of the initial seed particles that gave a high yield of electron-positron pairs. At the end of the simulation, the above-mentioned ten initial particles originated approximately $1.64\times10^6$ electron-positron pairs and approximately $1.65\times10^8$ photons with energy larger than 2.5~MeV in the collision of two laser pulses with 300~PW power, four wavelength waist radius and 20~fs duration. By contrast, by increasing the laser duration from 20~fs to 40~fs with all the other parameters unchanged, all particles are expelled out of the focal volume well before the laser pulse peaks reach the focus. Figure~\ref{fig:2} shows the trajectories of four of the above-mentioned ultraprolific initial particles driven by two 300~PW laser pulses both for $\tau=40$~fs duration (Fig.~\ref{fig:2}(a)-(b)) and for $\tau=20$~fs duration (Fig.~\ref{fig:2}(c)-(d)). While for 40~fs laser pulses all particles are expelled out of the focal region when the intensity at the focus is $I \approx 9.3 \times 10^{19}~\text{ W/cm$^2$}$, for 20~fs laser pulse these ultraprolific initial particles exhibit a complex dynamics, and remain trapped into the focal region for the whole laser pulse duration (see Fig.~\ref{fig:2}). Figure~\ref{fig:3} displays the details of the complex trajectory of one of the ultraprolific initial particles and its projection on the $xy$, $xz$ and $yz$ planes. The particle undergoes rapid changes of trajectory within a scale noticeably smaller than the laser wavelength. Note that the particle's dynamics is also considerably affected by single photon emissions.
In addition, we stress that when the number of particles triggering a cascade is small, then large statistical fluctuations are possible in the electron-positron yield. This has been already shown, for example, in Fig.~3 of the Supplementary information of Ref.~\cite{tamburiniSR17}, where the different results of the final electron-positron yield obtained from 100 different runs of the same QED cascade initiated by a single particle are reported.

The effect of the laser pulse duration $\tau$ on the number and energy of the produced photons is illustrated in Fig.~\ref{fig:4}. Figure~\ref{fig:4}(a) shows that for 20~fs laser pulses high energy photons with several hundreds MeV are generated as the peaks of the laser pulses collide, whereas for 40~fs only few photons with energy below 1~MeV are generated, as shown in Fig.~\ref{fig:3}(b). In addition, for 40~fs laser pulses all photon emissions occur in a relatively short temporal window, which is noticeably before the peaks of the two laser pulses overlap. The abrupt fall in the number of photon emissions long before the peaks of the laser pulses reach the focus confirms that all seed electrons have escaped the focal volume (see Fig.~\ref{fig:4}(b)). 

Finally, we discuss the effect on a seeded QED cascade of either a relative delay $\Delta t$ or a focal axis misalignment $\Delta x$ between the two laser pulses. In the case of the presence of a temporal offset $\Delta t$, the collision between the two laser pulses occurs at $z^*=c\Delta t /2$ and the relevant quantity is the Rayleigh length $z_{\text{R}} = \pi w_0^2 /\lambda$ of the laser pulses, where $w_0$ is the laser waist radius. In fact, after propagating for a distance $z_{\text{R}}$ the laser pulse radius $w(z)=w_0\sqrt{1+(z/z_R)^2}$ (spot area) increases by a factor $\sqrt{2}$ (2) and therefore the laser field (laser intensity) decreases by a factor $\sqrt{2}$ (2). Thus, it is expected that the laser pulse is weakly affected by diffraction when propagating over distances much smaller than $z_R$, whereas the effect of $\Delta t$ is expected to be important when $z^* \gtrsim z_R$, which implies $\Delta t \gtrsim 2 \pi w_0^2/(\lambda c)$, i.e. $\Delta t \approx268\text{ fs}$ with $w_0=4\lambda$. Note that a delay $\Delta t \approx268\text{ fs}$ is much larger than the duration of the laser pulses considered here $\tau$=10-40~fs. Similarly, due to the rapid decay of the laser fields outside the laser pulse waist radius $w_0$, a significant reduction of the electron-positron yield is expected for a focal axis misalignment $\Delta x$ comparable or larger than $w_0$.

The simulation results obtained as a function either of $\Delta t$ or of $\Delta x$ are summarized in Tab.~\ref{tab:2} and Figs.~\ref{fig:tab}(b)-\ref{fig:tab}(c). In agreement with the above-mentioned qualitative expectations, Tab.~\ref{tab:2} shows that the development of a QED cascade is weakly affected by a delay $\Delta t$, as the decrease in the electron positron pair yield is appreciable only when the collision of the two pulses occurs at one Rayleigh length from the focus. On the other hand, the presence of a $\Delta x \gtrsim w_0$ results into a dramatic decrease in the number produced particles (see Tab.~\ref{tab:2}).
Qualitatively, this can be understood by assuming that the electron dynamics is dominated only by one of the two laser pulses. If $\Delta t > \tau$ and $\Delta x > w_0$, this approximation is roughly applicable at least at the beginning of the interaction. This corresponds to consider only the first laser pulse that reaches the focus in the case of a delay $\Delta t$ (see Fig.~\ref{fig:1}(b)), and to consider only the laser pulse with focal axis closer to the electron position in the case of a misalignment $\Delta x$ (see Fig.~\ref{fig:1}(c)). 
Since hydrogen ionization occurs at intensities of the order of $10^{14}\text{ W/cm$^2$}$, the electron is approximately at rest, initially. In fact, an intensity $I_0 \approx 10^{14}\text{ W/cm$^2$}$ corresponds to $\bm{a} (\varphi_0, r_0) \approx 10^{-2} m_e c$, and the transverse and longitudinal components of the momentum within a temporal interval such that $|r(t)-r_0| \ll w_0$ are roughly given by $\bm{p}_\perp (\varphi, r_0) \approx  m_e c \bm{a} (\varphi, r_0)$ and $p_\parallel (\varphi, r_0) \approx  m_e c \bm{a}^2 (\varphi, r_0) /2$, respectively. Here $\varphi = \omega (t - z(t)/c)$ is the laser phase at the electron position $r=\sqrt{x^2(t)+y^2(t)}$ is the electron distance from the focal axis in its orthogonal plane, and $r_0$ is its initial value. Thus, electrons oscillate along the polarization axis due to the laser electric field and oscillate and are accelerated along the laser propagation direction by the $\bm{v}\times\bm{B}$ force. In the presence of a delay $\Delta t$ electrons are therefore slightly accelerated towards the second laser pulse, and subsequently interact with both laser pulses simultaneously. This does not significantly alter the development of a QED cascade unless $\Delta t$ is large enough for electrons to move in the $xy$ plane for a length comparable to $w_0$ before the second pulse reaches the focus. Note that, when electrons are simultaneously driven by both laser pulses, then the dynamics due to the $\bm{v}\times\bm{B}$ force is much more complex and can no longer be described by a regular ponderomotive potential~\cite{bauerPRL95}. By contrast, when the focal axis of the two pulses are misaligned with $\Delta x \gtrsim w_0$, most of the electrons basically never interact strongly with the second laser pulse, but are accelerated far from the focal region by the first laser pulse. In this case the electron quantum parameter $\chi_e$ remains smaller than unity for most of the electrons, and the probability of generation of $e^-e^+$ pairs is therefore exponentially suppressed. It is worth mentioning that statistical fluctuations associated with the small number of initial particles $N_{0,e^{-}}$ actually seeding the production of electron positron pairs in some simulations resulted in a larger final electron-positron yield in the presence of a 20~fs relative delay than without any relative delay (see Tab.~\ref{tab:2}).

\section{Summary and Conclusions}

In summary, we have investigated with numerical simulations the effect on seeded QED cascades of the laser pulse duration, the presence of a relative time delay, and the presence of a focal axis misalignment in the head-on collision of two laser pulses. We have shown that, similarly to the laser pulse waist radius, the laser pulse duration can be employed as a control parameter to facilitate or prevent the onset of seeded QED cascades. In fact, for a fixed waist radius the number of seed particles expelled out of the focal volume increases with increasing laser pulse duration. This occurs because the time available to seed particles to exit the focal volume before initiating a cascade increases for longer laser pulses. Thus, the optimal development of a QED cascade requires a relatively large waist radius for longer pulses, whereas short laser pulses can trigger a cascade with tighter focusing. In addition, we have shown that while a relative delay $\Delta t \ll 2 \pi w_0^2 / (\lambda c)$ does not significantly affect a QED cascade, the presence of a transverse misalignment of the laser focal axis $\Delta x$ comparable to the waist radius $w_0$ dramatically suppresses the generation of high energy photons and of electron-positron pairs.

\begin{acknowledgements}
We would like to thank Prof.~C.~H.~Keitel for his comments and his appreciation of this work.
\end{acknowledgements}


%
%

\end{document}